\documentclass[prd,12pt,aps,superscriptaddress,preprintnumbers,showpacs,tightenlines,floatfix,nofootinbib,amssymb]{revtex4}
\usepackage{epsfig}
\usepackage{graphicx,color}

\newcommand{\beq}{\begin{equation}}
\newcommand{\eeq}{\end{equation}}
\newcommand{\beqn}{\begin{eqnarray}}
\newcommand{\eeqn}{\end{eqnarray}}
\newcommand{\bea}[1]{\beq\begin{array}{#1}}
\newcommand{\eea}{\end{array}\eeq}
\newcommand{\eq}[1]{(\ref{#1})}

\newcommand{\Tr}{{\mathrm {Tr}}\,}

\newcommand{\Z}{Z\!\!\! Z}

\newcommand{\cC}{{\cal C}}

\newcommand{\Kanazawa}{\affiliation{Institute for Theoretical Physics,
Kanazawa University, Kanazawa 920-1192, Japan}}

\newcommand{\ITEP}{\affiliation{Institute of Theoretical and
Experimental Physics, B.Cheremushkinskaya 25, Moscow, 117259, Russia}}

\begin{document}

\title{Matter degrees of freedom and string breaking \\
in Abelian projected quenched SU(2) QCD}

\author{M.N.~Chernodub}\Kanazawa\ITEP
\author{Koichi~Hashimoto}\Kanazawa
\author{Tsuneo~Suzuki}\Kanazawa

\preprint{KANAZAWA/2003-13}
\preprint{ITEP-LAT-2003-09}

\begin{abstract}
In the Abelian projection the Yang--Mills theory
contains Abelian gauge fields (diagonal degrees of freedom) and
the Abelian matter fields (off-diagonal degrees) described
by a complicated action. The matter fields are essential for the
breaking of the adjoint string. We obtain numerically the effective action
of the Abelian gauge and the Abelian matter fields in quenched $SU(2)$ QCD and
show that the Abelian matter fields provide
an essential contribution to the total action even in the infrared region. We also
observe the breaking of an Abelian analog of the adjoint string using Abelian operators.
We show that the adjoint string tension is dominated by the Abelian and the monopole
contributions similarly to the case of the fundamental particles.
We conclude that the adjoint string breaking can successfully be
described in the Abelian projection formalism.
\end{abstract}


\date{November 17, 2003}

\maketitle

\section{Introduction}
\label{one}

The mechanism of color confinement in QCD is one of the most important
non--perturbative problems in the quantum field theory.
One of the most promising approaches to this problem
is based on the existence of the dual objects, called monopoles, which
are condensed in the confinement phase. This approach
-- known as the dual superconductor hypothesis~\cite{DualSuperconductor} --
is realized with the help of the so called Abelian projection~\cite{'tHooft:1981ht}
of $SU(N)$ color degrees of freedom to $U(1)^{N-1}$ degrees of freedom.

The model was shown to be quite successful in explanation of the confinement
of the fundamental charges such as quarks (see, $e.g.$, reviews~\cite{Reviews}).
Abelian and monopole contributions
to the inter--quark potential are dominant in the long-range region of
quenched QCD~\cite{suzuki90,ref:bali}. An infrared effective monopole action has
been derived in the continuum limit after a block--spin
transformation of monopole currents~\cite{shiba95,nakamura}. It is
a quantum perfect action described by the monopole currents.
The condensation of the monopoles in the confinement phase was
observed in various numerical approaches~\cite{shiba95,MonopoleCondensation}.
In the language of the monopole currents the condensation implies the formation
of the percolating cluster studied both numerically~\cite{ref:MIP:clusters}
and analytically~\cite{ref:Zakharov}.

However, this mechanism has a serious problem even in quenched QCD.
Although the 't~Hooft scenario describes the confinement of
quarks correctly, this scenario predicts also the existence of the string tension for
the adjoint charges (gluons) in the infrared region. On the other hand gluon
charges must be screened at large distances due to the presence of the gluons in
the QCD vacuum. This screening--confinement problem was extensively discussed in
recent publications~\cite{StringBreaking}.

The problem of the screening of the adjoint charges in quenched $SU(N)$ QCD has also
been discussed in Ref.~\cite{greensite}. The paper provides arguments that the
relevant quantity in the confinement mechanism is not the Abelian
monopoles but the $Z(N)$ center-vortices which can explain the
screening problem~\cite{ZNpicture}. In our study we pursue a different approach
based on the dual superconductor model.

Consider the screening in a confining Abelian model with the charge--two
matter fields (take, for example, the Abelian Higgs model with the compact
gauge fields). The presence of doubly charged matter fields screens
the confining interaction between the external particles
with opposite double charges. This happens due to the pair creation from
the vacuum at certain separations between the external charges. As a
result, the potential between the particles flattens at some distances.
It should be stressed that the problem is not only to explain the flattening
of the potential but also to show the linear behaviour of the potential
in the intermediate region. On the other hand the charge--one external fields
remain unscreened in this model. Namely, the potential is linearly rising
at large distances.

The standard model of the dual superconductor in quenched QCD
ignores the existence of the off-diagonal gluons. However, these gluons
have a charge two with respect to the Abelian subgroup and they may explain
the flattening of the inter--gluon potential which is usually studied with the help
of the adjoint Wilson loop. On the other hand the introduction of the new degrees of freedom
-- the off-diagonal gluons -- should not violate already achieved success of the
explanation of the quark confinement in this model. Indeed, quarks have
the charge one and doubly charged gluons can not screen them\footnote{However, we may
expect a renormalization of the tension of the string spanned between the quarks due to
the presence of the double charges}. These and related issues were discussed
in Ref.~\cite{SuzukiChernodub02} for quenched as well as for full $SU(N)$ QCD.

From the point of view of a realization of the (modified) dual superconductor
scenario it seems that we have to keep all charge two Abelian Wilson
loops in the effective action written by the Abelian link fields to
reproduce the screening of charge two. Indeed, the theory in terms of the Abelian link
fields or the Abelian monopole currents alone becomes highly non--local if we integrate
out all off-diagonal gluon fields after an Abelian projection. Needless to say, such an
Abelian effective action is useless. The same problem is more serious in the real full QCD,
since a fundamental charge is also screened in this case.

The aim of this paper is to calculate numerically the effective action of quenched QCD
within the Abelian projection formalism. Contrary to previous calculations of this kind we include
also the doubly charged off-diagonal gluon fields into the effective action and we show that their
contribution is essential and thus can not be neglected. We also calculate correlators of the adjoint
Polyakov loops in the Abelian formalism and observe the screening of a properly defined potential
between static adjoint sources.

The plan of the paper is the following. In Section~\ref{sec:review} we discuss
how the screening and confinement problem is solved qualitatively in the framework of
Abelian dynamics. Section~\ref{sec:action} is devoted to the investigation
of the Abelian action for the Abelian gauge and matter fields obtained by the inverse
Monte-Carlo method. In Section~\ref{sec:breaking} we discuss the potential between the adjoint
($Q=2$) charges within the Abelian projection formalism. We show numerically that properly defined
Abelian potential shows screening of the $Q=2$ charges. Moreover, we observe the Abelian and
the monopole dominance for the adjoint string tension. Our conclusions are presented in the last Section.

\newpage

\section{String Breaking in Abelian Projected Theory}
\label{sec:review}

The partition function of the Abelian effective theory of quenched $SU(2)$ QCD
in the infrared region may be approximated in the Villain--like
form~\cite{chernodub}:
\beqn
Z_Q[J]=\int\limits_{-\pi}^{\pi}{\cal D}\theta\sum_{n\in\Z(c_2)}
e^{-\frac{1}{4\pi^2}((d\theta+2\pi n),\Delta D(d\theta+2\pi n))
+iQ(\theta,J)}\,,
\label{eq:ZJ}
\eeqn
where $D$ is a differential operator
\beqn
D\approx\bar{\alpha}+\bar{\beta}\Delta^{-1}+\bar{\gamma}\Delta\,.
\eeqn
This operator contains local self--interaction term, the
Coulomb term described by the inverse Laplacian, $\Delta^{-1}$, and
additional interactions between nearest neighbors. The coupling
constants $\bar{\alpha}$, $\bar{\beta}$ and $\bar{\gamma}$ were calculated
numerically in Ref.~\cite{chernodub}. To simplify notations we use the
differential form formalism on the lattice~\cite{DiffLattice}.

The partition function~\eq{eq:ZJ} can be rewritten as a string model~\cite{chernodub},
\beqn
Z_Q[J] \propto \sum_{\stackrel{\sigma\in\Z(c_2)}{\delta \sigma=QJ}}
e^{-\pi^2(\sigma,(\Delta D)^{-1}\sigma)}\,,
\label{eq:ZJ:string}
\eeqn
where we have neglected perimeter terms. This model does not contain
the dynamical matter fields and therefore the string variable $\sigma$
is always closed on the external current $J$. Therefore there is no source
for the string breaking in this model.

Now let us consider the off-diagonal gluons. The Wilson action
of quenched $SU(2)$ QCD is
\beqn
S = \frac{\beta}{2} \sum_{s,\mu,\nu} \, \Tr U_{\mu\nu}(s)\,,\quad
U_{\mu\nu}(s) = U_{\mu}(s)U_{\nu}(s+\hat{\mu})
U_{\mu}^{\dag}(s+\hat{\nu})U_{\nu}^{\dag}(s)\,,
\label{eq:action}
\eeqn
where $U_\mu(s)$ is the $SU(2)$ gauge field.

It is convenient to parameterize the $SU(2)$ link variable $U_{\mu}(s)$ as
$U_{\mu}(s) = c_{\mu}(s)u_{\mu}(s)$ where
\beqn
c_\mu(s) = \left(
\begin{array}{@{}cc@{}}
\cos\phi_{\mu}(s)&
i\sin\phi_{\mu}(s)
e^{-i\varphi_{\mu}(s)}\\
i\sin\phi_{\mu}(s)
e^{i\varphi_{\mu}(s)} &
\cos\phi_{\mu}(s)
\end{array} \right) \,,
\quad
u_\mu(s) = \left(
\begin{array}{@{}cc@{}}
e^{i\theta_{\mu}(s)} & 0 \nonumber\\
0 & e^{-i\theta_{\mu}(s)}
\end{array} \right)\,.
\label{eq:definitions:c:s}
\eeqn
Here $\theta$, $\varphi$ and $\phi$ are independent variables
defined in the regions $-\pi\le\theta_{\mu}(s), \varphi_{\mu}(s)<\pi$ and
$0\le\phi_{\mu}(s)<\frac{\pi}{2}$. The field $\theta$ behaves as a $U(1)$
gauge field while the field $\varphi$ corresponds to the phase of the
off--diagonal gluon field because under an Abelian gauge transformation,
\beqn
\Omega^{\mathrm{Abel}}(s) = {\mathrm{diag} (e^{i \alpha(s)},e^{- i \alpha(s)})}\,,
\label{eq:gauge:abelian}
\eeqn
they behave as follows:
\beqn
\theta_{\mu}(s)\to\theta_{\mu}(s)-\partial_{\mu}\alpha(s) \equiv
\theta_{\mu}(s)+\alpha(s)-\alpha(s+\hat{\mu})\,, \quad
\varphi_{\mu}(s)\to\varphi_{\mu}(s)+2\alpha(s)\,.
\eeqn

The variable $\phi_\mu(s)$ is not affected by the $U(1)$ gauge transformation.
After an Abelian projection we can integrate this variable out without
harming the $U(1)$ content of the model.
In order to get an insight of possible forms of interactions between the Abelian gauge and Abelian matter
fields we replace the averages of $\cos\phi_{\mu}(s)$ and $\sin\phi_{\mu}(s)$ by their
mean values:
\beqn
\cos\phi_{\mu}(s) \to \langle \cos\phi_{\mu}(s) \rangle \equiv c\,, \quad
\sin\phi_{\mu}(s) \to \langle \sin\phi_{\mu}(s) \rangle \equiv s\,,
\label{eq:mean:field}
\eeqn
where $c$ and $s$ are functions of the $SU(2)$ coupling constant~$\beta$.

As the Abelian projection, we use the Maximal Abelian gauge which is defined
by a maximization of the functional,
\beqn
R = \frac{1}{2} \sum_{s,\mu}{\rm Tr}\Big(\sigma_3 \widetilde{U}_\mu(s)
\sigma_3 \widetilde{U}^{\dagger}_\mu(s)\Big) \equiv
\sum_{s,\mu} (2 \cos^2\phi_{\mu}(s) -1 )\,,
\label{eq:R}
\eeqn
with respect to the $SU(2)$ gauge transformations,
$U_{\mu}(s) \to \widetilde{U}_\mu(s)=\Omega(s)U_\mu(s)\Omega^\dagger(s+\hat\mu)$.
The functional \eq{eq:R} is invariant under residual $U(1)$ gauge
transformations~\eq{eq:gauge:abelian}.
The local condition corresponding to maximization~\eq{eq:R} can be written in the continuum
limit as the differential equation $(\partial_{\mu}+igA_{\mu}^3)(A_{\mu}^1-iA_{\mu}^2)=0$.

The maximization of the functional~\eq{eq:R} corresponds to the minimization of the $\phi$
variable. Thus the observation of Refs.~\cite{suzukiNPBPS,Chernodub:pw} made for the mean
values~\eq{eq:mean:field},
\beqn
c \simeq 1\,, \quad s\ll c\,.
\label{eq:c:s:relations}
\eeqn
does not come as a surprise. These relations hold in a wide region of the coupling
constant~$\beta$.

Following Ref.~\cite{Chernodub:pw} we rewrite the action of the model~\eq{eq:action}
in terms of the variables~$\theta$, $\varphi$ and $\phi$ with the help of the
definitions \eq{eq:definitions:c:s}. Applying
Eq.~\eq{eq:mean:field} to the original action we get
\beqn
& & \frac{1}{2}\Tr U_{\mu\nu}(s) = c^4\cos(\Theta_{\mu\nu}(s))
-c^2s^2\cos(\Theta_{\mu\nu}(s)-H_{\mu\nu}(s)-C_{\mu\nu}(s)) \nonumber\\
&& - c^2s^2\cos(\Theta_{\mu\nu}(s)+H_{\nu\mu}(s)-C_{\mu\nu}(s))
+c^2s^2\cos(\Theta_{\mu\nu}(s)+H_{\nu\mu}(s))
\label{eq:action:expand}\\
&& +c^2s^2\cos(\Theta_{\mu\nu}(s)-H_{\mu\nu}(s) +H_{\nu\mu}(s)-C_{\mu\nu}(s))
+c^2s^2\cos(\Theta_{\mu\nu}(s)-H_{\mu\nu}(s))\nonumber\\
&&+c^2s^2\cos(\Theta_{\mu\nu}(s)+C_{\mu\nu}(s))
+s^4\cos(\Theta_{\mu\nu}(s)-H_{\mu\nu}(s) +H_{\nu\mu}(s)-2C_{\mu\nu}(s)), \nonumber
\eeqn
where we have denoted the $U(1)$ gauge invariant variables as follows:
\begin{eqnarray}
\Theta_{\mu\nu}(s) & = &\theta_{\mu}(s)+\theta_{\nu}(s+\hat{\mu})
-\theta_{\mu}(s+\hat{\nu})-\theta_{\nu}(s)\,,
\label{eq:Theta}\\
H_{\mu\nu}(s) & = &2\theta_{\mu}(s)+\varphi_{\nu}(s)
-\varphi_{\nu}(s+\hat{\mu})\,,
\label{eq:H}\\
C_{\mu\nu}(s) & = & \varphi_{\mu}(s)-\varphi_{\nu}(s)\,.
\label{eq:C}
\end{eqnarray}
The variable $\Theta$ is the $U(1)$ plaquette for the gauge field $\theta$, the variable
$H$ describes the interaction of the matter field $\varphi$ with the gauge field $\theta$
and the variable $C$ corresponds to the self--interaction of the matter field. The validity
of the mean field approximation based on a self--consistent substitution~\eq{eq:mean:field}
is not known. When we perform the $\phi$ integration, we generally get an effective action in
terms of $\Theta_{\mu\nu}$, $H_{\mu\nu}$ and $C_{\mu\nu}$. Below we use numerical method to find
this effective action.

A few remarks about the action~\eq{eq:action:expand} are now in order.
(i) {}From~Eq.~\eq{eq:c:s:relations} one can immediately observe that the leading contribution
to the action is provided by the first QED--like term depending on the variables $\theta$ only.
The coupling between the gauge field $\theta$ and the matter field $\varphi$ is suppressed
and the self--interaction of the matter field is suppressed even further. (ii) The
action~\eq{eq:action:expand} should acquire corrections from the Faddeev--Popov
determinant resulting from the fixing of the Maximal Abelian gauge. This determinant is an essentially
non--local functional and the leading local terms were calculated in Ref.\cite{Chernodub:pw}.

Let us assume for simplicity the following effective action:
\beqn
S_{{\rm eff.}} = S^{(1)}(\theta)+S^{(2)}(\theta,\varphi)\,,\quad
S^{(2)}(\theta,\varphi)&=&-F_1(H)-F_2(H')-F_3(C)\,.
\label{eq:action:effective}
\eeqn
where we put $H=H_{\mu\nu}(s)$, $H'=H_{\nu\mu}(s)$, $C=C_{\mu\nu}(s)$
and $F_1$, $F_2$, $F_3$ are periodic functions.
Following Ref.~\cite{SuzukiChernodub02} we rewrite the corresponding
partition function $Z$ with the external source $J$ as follows:
\beqn
Z_Q[J]&=&\int\limits_{-\pi}^{\pi} {\cal D}\theta {\cal D}\varphi \,
e^{-S_{{\rm eff.}}+iQ(\theta,J)} = \int\limits_{-\pi}^{\pi} {\cal D}\theta {\cal D}\varphi
\,\, e^{-S^{(1)}(\theta) -S^{(2)}(\theta,\varphi)+iQ(\theta,J)}\nonumber\\
&=&\int\limits_{-\pi}^{\pi} {\cal D}\theta
\hspace{1mm}e^{-S^{(1)}(\theta)+iQ(\theta,J)}
\left[\int\limits_{-\pi}^{\pi} {\cal D}\varphi \, e^{F_1(H)+F_2(H')+F_3(C)}\right]\,.
\label{eq:Z:J:effective}
\eeqn
The part in the square brackets can be expanded in the Fourier series,
\beqn
[\cdots] &=&\int\limits_{-\pi}^{\pi} {\cal D}\varphi\sum_{\stackrel{n^{(i)}\in \Z(c_2)}{i=1,2,3}}
I_1(n^{(1)})\, I_2(n^{(2)})\, I_3(n^{(3)})\, e^{i(H,n^{(1)})+i(H',n^{(2)})+i(C,n^{(3)})}\,.
\label{eq:brackets}
\eeqn
where $n^{(i)}$, $i=1,2,3$ are integers and the lattice tensors
$H$, $H'$, $n^{(1)}$, $n^{(2)}$ sum only for $\mu>\nu$ because $H$, $H'$ are not
anti--symmetric contrary to $(C,n^{(3)})$.

Integrating over $\varphi$ and summing over $n^{(3)}$ one can
rewrite Eq.~\eq{eq:brackets} as follows:
\beqn
[\cdots] =\sum_{\stackrel{j\in\Z(c_1)}{\delta j=0}} w(j)\, e^{2i(\theta,j)}\,,
\label{eq:brackets2}
\eeqn
where $w(j)$ are certain weights for the closed current $j$ which is defined from
the variables $n^{(1)}$ and $n^{(2)}$.
\beqn
j_{\mu}(s)=\sum_{\nu(<\mu)}n_{\mu\nu}^{(1)}(s) +\sum_{\nu(>\mu)}n^{(2)}_{\nu\mu}(s)\,.
\label{eq:j:text}
\eeqn
The general form of Eqs.~\eq{eq:brackets2} and \eq{eq:j:text}
follows from the fact that the fields $\varphi$ are doubly charged and from the
gauge invariance of the expression under the exponential function in Eq.~\eq{eq:brackets}.
We also give a detailed derivation of Eqs.~\eq{eq:brackets2} and \eq{eq:j:text} in
Appendix~\ref{app:brackets}.

To simplify further considerations let us rewrite the first term in Eq.~\eq{eq:action:effective}
in the Villain form as in Eq.~\eq{eq:ZJ}. Then we get for the partition function~\eq{eq:Z:J:effective}:
\beqn
Z_Q[J] \! = \! \int\limits_{-\pi}^{\pi} \! {\cal D}\theta \!\!\!\sum_{n\in\Z(c_2)}
\sum_{\stackrel{j\in\Z(c_1)}{\delta j=0}}\!
w(j)\, \exp\Bigl\{-\frac{1}{4\pi^2}((d\theta+2\pi n),
\Delta D(d\theta+2\pi n))+i(\theta,2j+QJ)\Bigr\}\,. \nonumber
\eeqn
Analogously to Eq.~\eq{eq:ZJ:string} we get the following model for the string variables
dual to the gauge field $\theta$:
\beqn
Z_Q[J]=\sum_{\stackrel{j\in{\Z}(c_1)}{\delta j=0}}
\sum_{\stackrel{\sigma\in{\Z}(c_2)}{\delta \sigma=2j+QJ}}
w(j) \,\exp\Bigl\{-\pi^2(\sigma,(\Delta D)^{-1}\sigma)\Bigr\}\,.
\label{eq:ZJ:string:effective}
\eeqn

The string model~\eq{eq:ZJ:string:effective} is different from the
model~\eq{eq:ZJ:string} by the presence of the doubly
charged currents representing the contribution of the off-diagonal gluons
(the first sum in Eq.~\eq{eq:ZJ:string:effective}). The second sum in this equation is
over the integer valued string variable which has the dynamical current $j$ as its boundary.

If the external charge has a unit value, $Q=1$, then the dynamical current $j$ can not
screen the external current $QJ$ and therefore the string always spans on the trajectories
of the external currents, $\delta \sigma = 2j+QJ \neq 0$. However, if the external current
is doubly charged, $Q=2$, then there exists the dynamical current $j=-J$ such
that $\delta\sigma=0$. This state breaks the string: when the distance between the
external charges is large enough the state with $j=-J$ provides a dominant contribution
to the partition function.

\section{Effective Action for Gauge and Matter fields}
\label{sec:action}

In this Section we calculate numerically the effective action for the Abelian gauge
and the matter fields in quenched $SU(2)$ QCD. We have chosen a trial action
in the form:
\beqn
S_{\mathrm{eff}}(\theta,\varphi) &=&\alpha_1 S_1(\theta) +\alpha_2 S_2(\theta)
+ \alpha_3 S_3(\theta) + \beta_1 S_4(\theta,\varphi)\,,
\label{eq:Seff:trial}
\eeqn
where $\alpha_{i}$, $i=1,2,3$ and $\beta_1$ are the coupling constants to be determined
numerically.

The functionals $S_i$, $i=1,2,3$ describe the action of the gauge field $\theta$,
\beqn
S_1 &=&-\sum_{s,\mu\ne\nu}[\cos\Theta_{\mu\nu}(s)]\,,
\label{eq:s1}\\
S_2 & = & -\sum_{s,\mu\ne\nu}[\cos2\Theta_{\mu\nu}(s)]\,,
\label{eq:s2}\\
S_3&=&+\sum_{s,\mu\ne\nu}[\sin\Theta_{\mu\nu}(s) \sin\Theta_{\mu\nu}(s+\hat{\mu})]\,,
\label{eq:s3}
\eeqn
where the plaquette variable $\Theta$ is given in Eq.~\eq{eq:Theta}.
The action $S_1$ is the leading term in the Abelian action~\eq{eq:action:expand} corresponding
to quenched $SU(2)$ QCD in the mean--field approximation.
The parts $S_{2,3}$ are also included because they may arise naturally from the integration
over $\phi$.

As an interaction term between the gauge, $\theta$, and the matter, $\varphi$, fields
we adopt for simplicity
\beqn
S_4&=&-\sum_{s,\mu\ne\nu} \Bigl[\cos(\Theta_{\mu\nu}(s)-H_{\mu\nu}(s))+
\cos(\Theta_{\mu\nu}(s)+H_{\nu\mu}(s))\Bigr]\,,
\label{eq:s4}
\end{eqnarray}
where the plaquette variable $H$ is given in Eq.~\eq{eq:H}. We have not included other terms
from Eq.~\eq{eq:action:expand} into the trial action because it turns out that the minimal
form of the action~\eq{eq:Seff:trial} describes the numerical data with a good accuracy.

\begin{figure}[h]
\includegraphics[angle=-0,scale=0.35,clip=true]{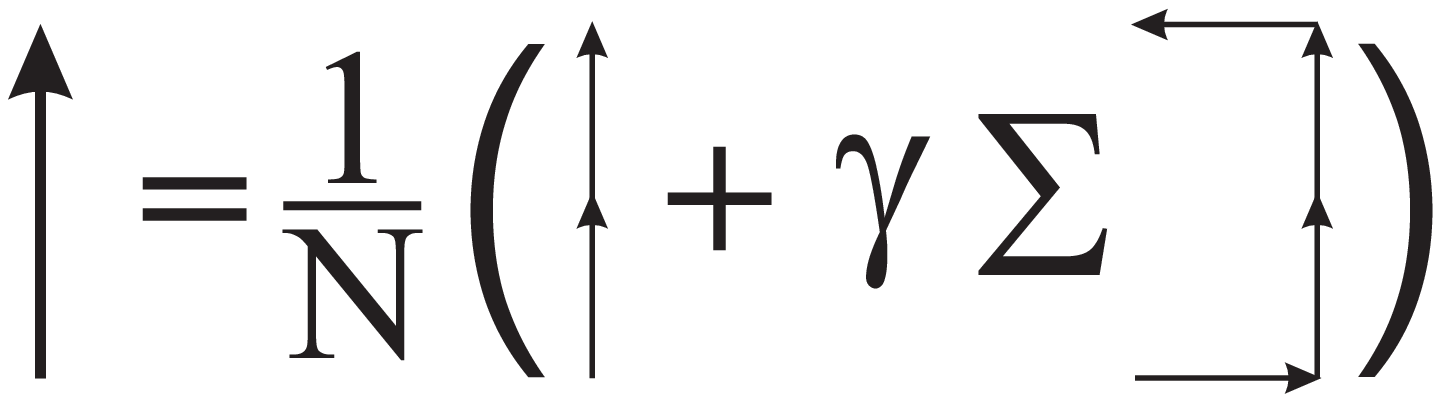}
\caption{The visualization of the blockspin transformation, Eq.~\eq{eq:block:spin}.}
\label{fig:blockspin}
\end{figure}

We have used the standard Monte--Carlo procedure to generate the gauge field configurations
on the $32^4$ lattice. The $SU(2)$ coupling constant was chosen in the range $\beta=2.1 \sim 2.7$.
We have generated 100 configurations of the gauge field for each value of the coupling constant
and then used the Simulated Annealing method~\cite{ref:bali} to fix the Maximal Abelian gauge.
The couplings $\alpha_i$, $i=1,2,3$ and $\beta_1$ were determined by solving the Schwinger--Dyson
equations~\cite{ref:Okawa}. We describe the details of this method in Appendix~\ref{app:Schwinger:Dyson}.
To make further improvement of our results towards the continuum limit
we used also a blockspin transformation for the $SU(2)$ link variable $U_{\mu}(s)$:
We apply the blockspin transformation to the link variable $U_{\mu}(s)$,
\beqn
U'_{\mu}(s')=\frac{1}{N}\left(U_{\mu}(s)U_{\mu}(s+\hat{\mu})
+\gamma\sum_{\nu(\ne\mu)}U_{\nu}(s)U_{\mu}(s+\hat{\nu})
U_{\mu}(s+\hat{\mu}+\hat{\nu})U^{\dag}_{\nu}(s+2\hat{\mu})
\right)\,.
\label{eq:block:spin}
\eeqn
which is visually represented in Figure~\ref{fig:blockspin}.
Here $N \equiv N(U)$ is the normalization factor which is introduced
to make the fat link belonging to the $SU(2)$ group.
The weight parameter $\gamma$ was set to $\gamma=0.5$.

\begin{figure}[!thb]
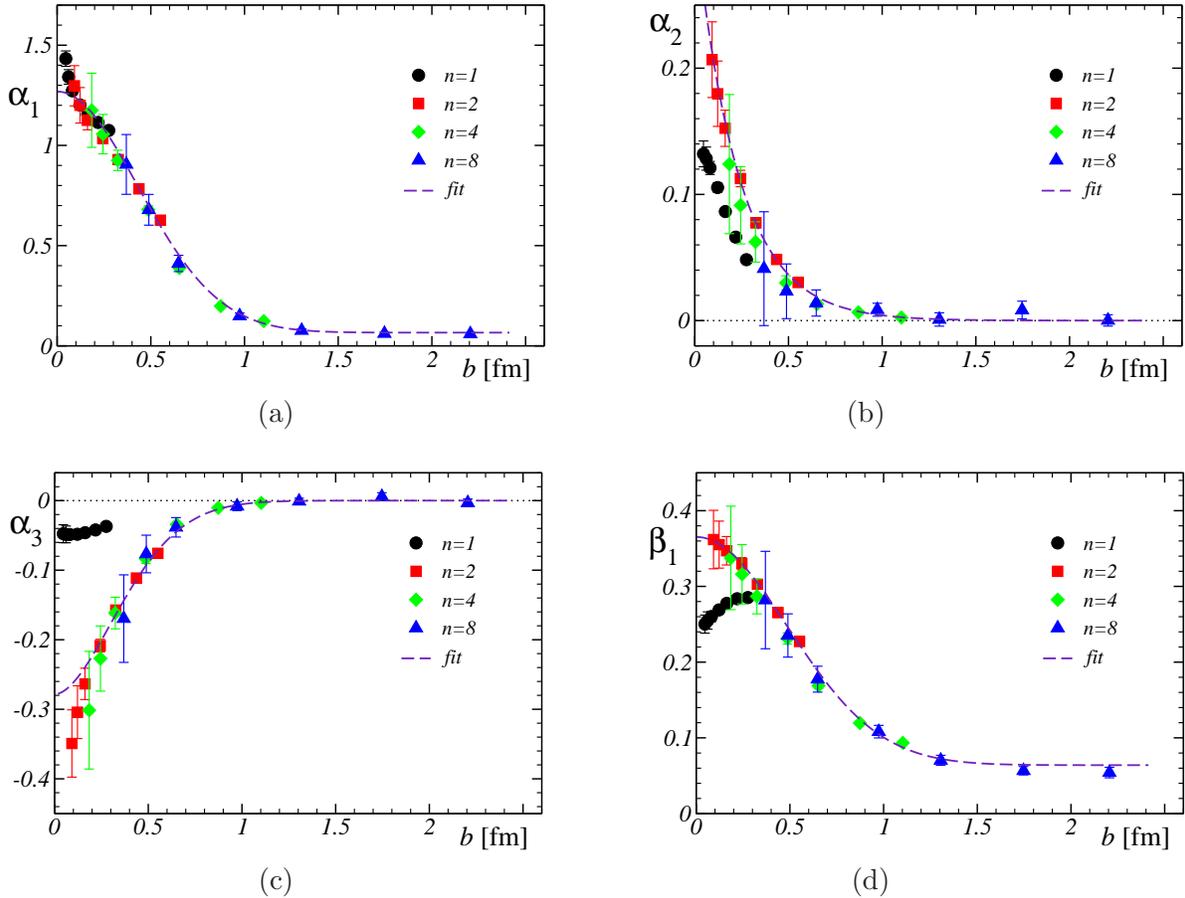

\begin{tabular}{cc}
\includegraphics[angle=-00,scale=0.3,clip=true]{alpha1.fit.eps}
&
\hskip 12mm\includegraphics[angle=-0,scale=0.3,clip=true]{alpha2.fit.eps}
\\
(a) & (b)
\vspace{5mm}
\\
\includegraphics[angle=-00,scale=0.3,clip=true]{alpha3.fit.eps}
&
\hskip 12mm\includegraphics[angle=-0,scale=0.3,clip=true]{beta1.fit.eps}
\\
(c) & (d) \\
\end{tabular}
\caption{The parameters $\alpha_i$, $i=1,2,3$ and $\beta_1$ for different blocking steps $n$
$vs.$ the scale parameter $b$. The fits by Eq.~\eq{eq:exponential:fit} are shown by the
dashed lines.}
\label{fig:alpha:beta}
\end{figure}
The couplings obtained in this way are depicted in Figures~\ref{fig:alpha:beta}(a-d).
The coupling $\alpha_1$ shows a perfect scaling since the coupling
constant depends only on the physical length $b$,
(and it does not depend on $n$ and $a$ separately). For the couplings $\alpha_2$, $\alpha_3$
and $\beta_1$ this feature does not work: the original data (no blockspin transformation, $n=1$)
is quite different from the cases where the blockspin transformation was done ($n>1$) while
the coupling constants with $n > 1$ scale almost perfectly.  One can make a conclusion
that the original data corresponds to very small values of $b$ where the effective action
takes more complicated form than~\eq{eq:Seff:trial}.

In order to quantitatively characterize the dependence of the coupling constants on
the scale factor~$b$ we have fitted the data by a function
\beqn
f(b) = C_0 + C_1 \exp\{- ({b \slash b_0)}^\nu\}\,,
\label{eq:exponential:fit}
\eeqn
where $C_{0,1}$, $\nu$ and $b_0$ are the fitting parameters. In our fits we have excluded
the data without the blockspin transformation, $n=1$, for all coupling constants except for $\alpha_1$
case. The best fit curves are
plotted in Figures~\ref{fig:alpha:beta} as the dashed lines, and the best fit parameters
are shown in Table~\ref{tbl:alpha:beta:fits}.
\begin{table}
\begin{tabular}{|c|c|c|c|c|}
\hline
coupling & $C_0$ & $C_1$ & $b_0$, [fm] & $\nu$ \\
\hline
$\alpha_1$ & 0.066(10) & 1.20(2) & 0.61(1) &2 \\
$\alpha_2$ & 0 & 0.32(2) & 0.231(7) & 1\\
$\alpha_3$ & 0 & -0.28(3) & 0.46(3) & 1.8(2) \\
$\beta_1$ & 0.064(5) & 0.30(1) & 0.69(2) & 2 \\
\hline
\end{tabular}
\caption{The parameters for the exponential fits~\eq{eq:exponential:fit}
of the couplings $\alpha_i$, $i=1,2,3$ and $\beta_1$.}
\label{tbl:alpha:beta:fits}
\end{table}

We have found that in the case of $\alpha_1$ and $\beta_1$ the parameter $\nu$ is very close to
two, and therefore in these fits we fixed this parameter, $\nu=2$. Similarly, we have also
fixed $\nu=1$ for $\alpha_2$ and $C_0=0$ for $\alpha_{2,3}$. Note that the fit
can not describe the coupling $\alpha_1$ accurately at small scales, $b \leqslant 0.2$~fm.
A similar deviation can be found for the coupling $\alpha_3$. We expect that at small scales
the Abelian action becomes much more complicated than the trial action
(\ref{eq:Seff:trial},\ref{eq:s1},\ref{eq:s2},\ref{eq:s3},\ref{eq:s4}) which we used
to solve the Schwinger--Dyson equations. The small similar effect is observed for the effective monopole
action obtained by the inverse Monte--Carlo methods~\cite{chernodub}.

The functional $S_1$, Eq.~\eq{eq:s1}, makes the leading contribution to the action since
the corresponding coupling, $\alpha_1$, is the largest.
The actions $S_2$ and $S_3$, in addition to the expected action $S_1$, play an essential role
at small scales since the corresponding couplings, $\alpha_2$ and $\alpha_3$, are non--vanishing.
The action $S_4$, which describes the interaction of the matter fields with the gauge fields, has
a non--vanishing coupling both at small and large scales similarly to $S_1$.
Moreover, according to Table~\ref{tbl:alpha:beta:fits} the
couplings $\alpha_1$ and $\beta_1$, corresponding to these parts of the total action, have
relatively large lengths $b_0$ compared to the coupling constants $\alpha_2$ and $\alpha_3$.
Thus, at large scales, $b \sqrt{\sigma} \gg 1$, the effective Abelian action for the $SU(2)$ gauge
theory can be approximated as a sum of the QED--like action for the gauge field, $S_1(\theta)$, and
the interaction term $S_4(\theta,\varphi)$.

We interpret the results, obtained in this Section as the manifestation of the
Abelian dominance (non--vanishing dominant coupling $\alpha_1$) and the importance of the off-diagonal (matter)
degrees of freedom (non--vanishing coupling $\beta_1$). The matter fields are essential for the
breaking of the adjoint string. {}From the point of view of further analytical study the results of this
Section are qualitative because in order to make a quantitative analytical predictions at
a finite value of the scale $b$ we need much more terms in the trial action~\eq{eq:Seff:trial}
than we have imposed. Indeed, in Ref.~\cite{chernodub} the monopole contribution to the string
tension has been calculated using the effective monopole action. The monopole action was obtained
numerically and it turns out that in order to get a correct analytical result for the string tension
one should take into account not only the most local terms in the effective monopole action but also
a series of the non--local terms. The situation with the effective action for the Abelian
fields~\eq{eq:Seff:trial} should be similar to the case of the monopole action since these actions
are related to each other~\cite{chernodub}. Nevertheless the adjoint string breaking can {\it quantitatively}
be discussed within the numerical approach on the basis of the Maximal Abelian gauge fixing.
This topic is discussed in the next Section.

\section{$Q=2$ Potential from Polyakov Loops}
\label{sec:breaking}

The easiest way to observe numerically the string breaking effect is to consider the theory
at finite temperature and define the potential with the help of the Polyakov loop
correlators~\cite{ref:PolyakovLoopOthers,SuzukiChernodub02}:
\begin{eqnarray}
\langle P(\vec{x})P^{\dag}(\vec{y})\rangle&=&e^{-V(\vec{x}-\vec{y})/T}\,.
\label{eq:V}
\end{eqnarray}
Here $T$ is temperature.

The adjoint Polyakov loop, $P_1$, is defined as follows:
\beqn
P_1 = \frac{1}{3} \Tr \left(\prod_{i\in \cC} D_1[U_i]\right)
= \frac{1}{3}(4p_0^2-1)\,,
\label{eq:w1}
\eeqn
where the color vector $p=p_0+i\vec{p}\cdot\vec{\sigma}$ defines the
fundamental Polyakov loop, $P_{1/2} = 1/2 \Tr p$, $p=\prod_{i\in \cC} U_i$,
and $\cC$ is the straight line parallel to the temperature direction.
The adjoint Polyakov loop~\eq{eq:w1} contains the charged term, $Q=2$, and
neutral term, $Q=0$:
\beqn
P_{Q=2} = \frac{2}{3}(p_0^2-p_3^2)\,, \quad P_{Q=0} = \frac{1}{3}(2 p_0^2+2 p_3^2-1)\,.
\eeqn

The Abelian dominance in the most general sense means that a non--Abelian observable
can be calculated with a good accuracy with the help of the corresponding
Abelian operator in a suitable Abelian projection. The Abelian dominance was first
established for the tension of the chromoelectric string spanned between the fundamental
sources~\cite{suzuki90}. In this case the non--Abelian Wilson (or, Polyakov) loop
was replaced by its Abelian counterpart.

However, in the case of the adjoint potential we
immediately encounter a problem~\cite{ref:greensite:caution}: in the Abelian
projection the $Q=2$ charged component of the Wilson loop shows the area law while the
neutral $Q=0$ component is constant. Therefore, strictly speaking, the straightforward
Abelian projection of the adjoint operators leads to the vanishing Abelian string tension.
The simplest way to overcome this difficulty is to introduce the obvious prescription
for the adjoint operators proposed originally in Ref.~\cite{ref:poulis}. Namely, one should
disregard the $Q=0$ component of the Wilson loop operator and
consider the $Q=2$ Abelian component of the Wilson loop as the Abelian analog of the
full (non-Abelian) loop. In Ref.~\cite{ref:poulis} some numerical arguments in favor of
the validity of this prescription were given. Below we follow this recipe and show
that the string breaking effect can indeed be seen in the $Q=2$ Abelian and monopole
components of the potential. Moreover, we have observed the Abelian and monopole dominance
for the adjoint string tension.

After the Abelian projection the $Q=2$ component becomes
\beqn
P_{Q=2}^{\rm ab}=\cos 2 \vartheta_\cC\,,\quad \vartheta_\cC = \sum_{i\in \cC}\theta_i\,.
\eeqn
where $\vartheta_\cC$ enters the $Q=1$ Abelian Polyakov loop,
$P^{\rm ab}_{1/2} = \cos \vartheta_\cC$.

We calculate numerically the static potential between the adjoint particles using
the Polyakov loop correlators~\eq{eq:V}. We use four types of the Polyakov loops:
non-Abelian, Abelian, monopole and photon Polyakov loops,
\beqn
P_{Q=2} = p_0^2-p_3^2\,,\quad
P_{Q=2}^{\rm ab} =  \cos2\vartheta_\cC\,,\quad
P_{Q=2}^{\rm mon} = \cos2\vartheta_\cC^{\rm mon}\,,\quad
P_{Q=2}^{\rm ph} = \cos2\vartheta_\cC^{\rm ph}\,,
\eeqn
respectively.

The functions $\vartheta_\cC^{\rm mon}$ and $\vartheta_\cC^{\rm ph}$ represent the contributions
to the Polyakov loop coming from the monopole currents and the photon fields,
respectively~\cite{ref:bali,suzuki90}:
\begin{eqnarray}
\vartheta_\cC^{\rm mon}&=&-\sum_t\sum_{\vec{x}',t'}D(\vec{x}-\vec{x}',t-t')
\partial'_{\mu}\bar{\Theta}_{\mu4}(\vec{x}',t')\,,\\
\vartheta_\cC^{\rm ph}&=&-2\pi\sum_t\sum_{\vec{x}',t'}D(\vec{x}-\vec{x}',t-t')
\partial'_{\mu}n_{\mu4}(\vec{x}',t')\,,
\end{eqnarray}
where the variables $\bar{\Theta} \in (-\pi,\pi)$ and $n \in \Z$ are extracted
from the Abelian plaquette
variable, $\Theta_{\mu\nu}(s) \equiv \theta_{\mu}(s) + \theta_{\nu}(s+\hat \mu)
- \theta_{\mu}(s+\hat \nu) - \theta_{\nu}(s) = \bar{\Theta}_{\mu\nu}(s)+2\pi n_{\mu\nu}(s)$.
$D(s)$ is the inverse Laplacian, $\partial'_{\mu}\partial_{\mu}D(s)=-\delta_{0,s}$.

We numerically measured the potential between the static adjoint sources on the $16^3\times 4$
lattice at $\beta=2.2$ (confinement phase) using 2000 configurations. The Abelian, monopole
and the photon components of the potential were measured in the Maximal Abelian gauge. In order
to reduce the statistical errors in our
calculations of the potentials we have applied the Hypercubic Blocking~\cite{ref:HCB} procedure
to ensembles of the non--Abelian, Abelian and photon gauge fields. We have not applied the blocking to the
monopole contribution of the potential because in this particular case the blocking makes the data noisier.
The Hypercubic Blocking method is briefly described in Appendix~\ref{app:Hypercubic:Blocking}.

\begin{figure}[!thb]
\includegraphics[angle=-00,scale=0.5,clip=true]{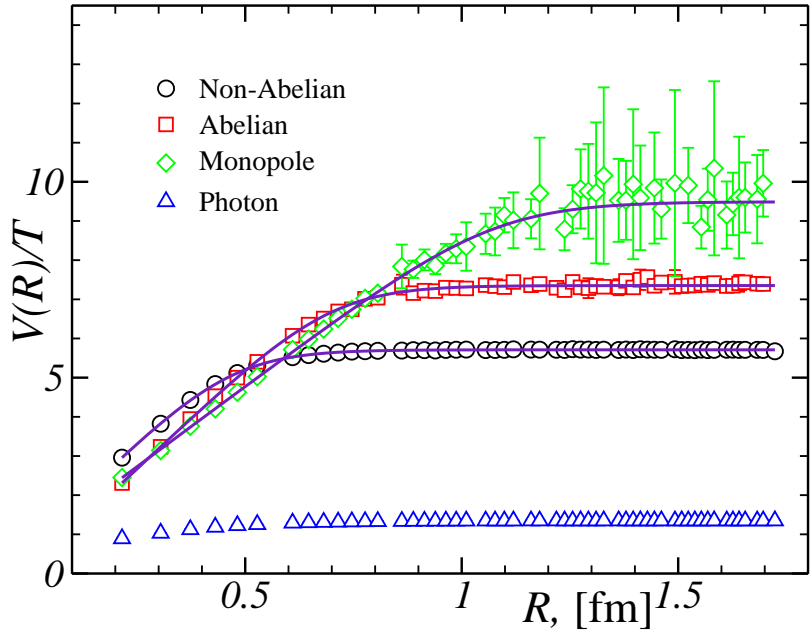}
\caption{The potential between adjoint static sources and the Abelian, the monopole and
the photon contributions to it. The fits by the function~\eq{eq:potential:fitting} are shown
by the solid lines.}
\label{fig:polyakov:loop:potential}
\end{figure}
We present the numerical results in Figure~\ref{fig:polyakov:loop:potential}. One can clearly
see that all potentials become flat in the infrared region clearly indicating the presence of the
string breaking. The non-Abelian potential as well as the Abelian and the monopole contributions
contain the linear pieces at small enough distances while the photon contribution to the potential
does not contain a linear part. These observations are in a qualitative agreement with the
Abelian (monopole) dominance hypothesis~\cite{suzuki90}.

To make a quantitative characterization of the potentials we fit our data by a function
\beqn
\exp\Bigl\{ - \frac{V^{\mathrm{fit}}(R)}{T}\Bigr\} =
\exp\Bigl\{- \frac{V_0 + 2 m}{T}\Bigr\} + \exp\Bigl\{- \frac{V_0 + V_{\mathrm{str}}(R)}{T}\Bigr\}\,,
\label{eq:potential:fitting}
\eeqn
where we have chosen the string potential in the simplest form, $V_{\mathrm{str}}(R) = \sigma_{Q=2} R$.
The fitting parameters are the adjoint string tension $\sigma_{Q=2}$, the mass parameter $m$
and the self--energy $V_0$. The first term in Eq.~\eq{eq:potential:fitting} corresponds to the
broken string state and the parameter $m$ is the mass of a state of
"external heavy adjoint source"--"light gluon". The second term is the unbroken string state. Here
we neglect other states including the string excitations.

We perform the fits in the range starting from two lattice spacings, $r_{\min} = 2 a$. The reason for
this restriction is twofold: (i) the hypercubic blocking modifies the potential at small distances;
(ii) in our fitting function~\eq{eq:potential:fitting} the perturbative Coulomb interaction (which
is essential at small distances) is not included\footnote{Nevertheless we have checked the effect
of the Coulomb interaction shifting the string potential as $V_{\mathrm{str}}(R) \to V_{\mathrm{str}}(R)
- \alpha/R$, where $\alpha$ is an additional fitting parameter. We have observed that the best fit
values of the parameters $\sigma_{Q=2}$ and $m$ got the shift about 1-2\% which is of the order
of the statistical errors for these parameters.}.

The best fit
functions are shown in Figure~\ref{fig:polyakov:loop:potential} by the dashed lines and
the best fit parameters are presented in Table~\ref{tbl:potential:fits}. One can clearly see
the existence of the Abelian dominance for the string tension:
$\sigma^{\mathrm{ab}}_{Q=2} \approx 0.94 \, \sigma_{Q=2}$, where $\sigma_{Q=2}$ is the string tension extracted
from the non-Abelian Polyakov loop correlator. The monopole dominance can also be observed:
$\sigma^{\mathrm{mon}}_{Q=2} \approx 0.83 \, \sigma^{\mathrm{ab}}_{Q=2} \approx 0.78 \, \sigma_{Q=2}$.
The monopole dominance is less manifest than the Abelian dominance in agreement with the precise
observations at $\beta = 2.5115$ in the case of fundamental external sources, Ref.~\cite{ref:bali}.

In Ref.~\cite{ref:bali} the potential between the static $Q=2$ Abelian sources has been measured in the zero
temperature case. Despite the string breaking has not been observed in this case, the ratio between $Q=2$
and $Q=1$ Abelian string has been measured: $\sigma_{Q=2}/\sigma_{Q=1} = 2.23(5)$. Taking into account that
the ratio between $Q=1$ Abelian and $SU(2)$ string tensions is~\cite{ref:bali},
$\sigma_{Q=2}/\sigma = 0.92(4)$, we get the prediction of Ref.~\cite{ref:bali} for the
ratio $\sigma_{Q=2}/\sigma = 2.42(12)$. We observe a very good agreement with our result,
$\sigma_{Q=2}/\sigma = 2.33(3)$, given in Table~\ref{tbl:potential:fits}.

\begin{table}
\begin{tabular}{|c|c|c|}
\hline
type & $\sigma_{Q=2}/\sigma$ & $m \slash \sqrt{\sigma}$\\
\hline
Non-Abelian & 2.49(3) & 1.28(1) \\
Abelian     & 2.33(3) & 1.84(1) \\
Monopole    & 1.94(1) & 2.27(2) \\
\hline
\end{tabular}
\caption{The parameters for the fits of the potential by the function~\eq{eq:potential:fitting}.
Here $\sigma \equiv \sigma_{1/2}(T=0)$.}
\label{tbl:potential:fits}
\end{table}

According to our numerical results the Abelian and the monopole contributions to the masses of
the heavy--light adjoint particles, $m$, do not coincide with the corresponding mass measured
with the help of the non-Abelian Polyakov loops. On the other hand we do not expect neither
Abelian nor monopole dominance to hold in this case since these types of dominance are usually
valid for infrared (non--local) quantities in accordance with the ideas of Ref.~\cite{DualSuperconductor}.
Because of the local nature of the mass $m$ the Abelian/monopole dominance may not work in this case.

The absence of the Abelian dominance for the mass parameter $m$ implies the absence of the
Abelian dominance for the string breaking distance. Indeed, the simplest definition of the string breaking
distance, $R_{\mathrm{sb}}$, corresponds to a value of $R$ at which both terms in Eq.~\eq{eq:potential:fitting}
are equal. For the linear string potential, $V_{\mathrm{str}} = \sigma_{Q=2} R$, this distance
is defined as $R_{\mathrm{sb}} = 2 m \slash \sigma_{Q=2}$. In other words, the string breaking
distance is the distance where the energy of the string, $\sigma_{Q=2} R_{\mathrm{sb}}$, is equivalent
to the energy of the two heavy--light states, $2 m$. Since the Abelian dominance works only
for the string tension $\sigma_{Q=2}$, the string
breaking distance, $R_{\mathrm{sb}}$, should not be Abelian/monopole dominated quantity.

\section{Conclusions}

We have calculated the effective action for the Abelian gauge and the Abelian charged matter fields
in the Maximal Abelian projection of quenched $SU(2)$ QCD. We have shown that
in the infrared limit the contribution of the matter field to the action is non--vanishing.
Thus we have shown at the qualitative level that the matter fields, carrying the Abelian charge
$Q=2$, must lead to the adjoint string breaking. To check this effect on the quantitative level
we have studied the potential between adjoint static sources as well as the Abelian and the
monopole contribution to this potential.
We have observed the string breaking (flattening of the adjoint potential) manifests itself
in the Abelian and the monopole contributions similarly to the non-Abelian case. Moreover,
we show that the adjoint string tension is dominated by the Abelian and the monopole
contributions analogously to the case of fundamental particles.
Thus we conclude that the adjoint string breaking can qualitatively be described in the
Abelian projection formalism. The key role in the adjoint string breaking in the Abelian
picture is played by the off-diagonal gluons which become the doubly charged Abelian vector
fields in the Abelian projection.

\appendix

\section{Derivation of Eq.~\eq{eq:brackets2}}
\label{app:brackets}

In this Appendix we present a detailed derivation of Eq.~\eq{eq:brackets2},
\beqn
\int\limits_{-\pi}^{\pi} {\cal D}\varphi \, e^{F_1(H)+F_2(H')+F_3(C)}
= \sum_{\stackrel{j\in{\Z}(c_1)}{\delta j=0}} w(j)\, e^{2i(\theta,j)}\,.
\label{eq:one}
\eeqn
The Fourier transformation applied to each of the terms in the {\it l.h.s.} of this equation gives:
\beqn
\int\limits_{-\pi}^{\pi} {\cal D}\varphi \, e^{F_1(H)+F_2(H')+F_3(C)}
=\int\limits_{-\pi}^{\pi} {\cal D}\varphi \prod_s \sum_{\stackrel{n_{\mu\nu}^{(i)}\in{\Z}}{i=1,2,3}}
\Bigl[\prod_{i=1}^3 I_i(n_{\mu\nu}^{(i)}(s))\Bigr]\, e^{i \Phi(\varphi, n^{(i)})}\,,
\label{eq:Fourier}
\eeqn
where $I_i$ are the Fourier components of $e^{F_i}$ and
$\Phi$ is the phase:
\beqn
\Phi(\varphi, n^{(i)}) = \sum_s\left[\sum_{\mu>\nu}(H_{\mu\nu}(s)n_{\mu\nu}^{(1)}(s)+
H_{\nu\mu}(s)n_{\mu\nu}^{(2)}(s))+ \sum_{\mu\ne\nu}C_{\mu\nu}(s)n_{\mu\nu}^{(3)}(s)\right]\,.
\eeqn
Using definitions \eq{eq:Theta}-\eq{eq:C} we get:
\beqn
& & \Phi(\varphi, n^{(i)}) =\sum_s\sum_{\mu>\nu}\Bigl\{
2\theta_{\mu}(s)n_{\mu\nu}^{(1)}(s) +2\theta_{\nu}(s)n_{\mu\nu}^{(2)}(s)
\label{eq:Phi}\\
&& +
\varphi_{\nu}(s) \Bigl[n_{\mu\nu}^{(1)}(s)-n_{\mu\nu}^{(1)}(s-\hat{\mu})-2n_{\mu\nu}^{(3)}(s)\Bigr]
+
\varphi_{\mu}(s) \Bigl[n_{\mu\nu}^{(2)}(s)-n_{\mu\nu}^{(2)}(s-\hat{\nu})+2n_{\mu\nu}^{(3)}(s)\Bigr]
\Bigr\}\,.
\nonumber
\eeqn
The integration over the field $\varphi$ gives two constraints
\beqn
n_{\mu\nu}^{(1)}(s)-n_{\mu\nu}^{(1)}(s-\hat{\mu})-2n_{\mu\nu}^{(3)}(s) = 0\,,
\qquad
n_{\mu\nu}^{(2)}(s)-n_{\mu\nu}^{(2)}(s-\hat{\nu})+2n_{\mu\nu}^{(3)}(s) = 0
\label{eq:constraints}
\eeqn
which lead to the
\beqn
n_{\mu\nu}^{(1)}(s)-n_{\mu\nu}^{(1)}(s-\hat{\mu}) +n_{\mu\nu}^{(2)}(s)-n_{\mu\nu}^{(2)}(s-\hat{\nu})=0\,.
\label{eq:constraint}
\eeqn

Equation \eq{eq:Phi} gives the natural definition for the current of the matter fields:
\beqn
j_{\mu}(s)=\sum_{\nu(<\mu)}n_{\mu\nu}^{(1)}(s) +\sum_{\nu(>\mu)}n^{(2)}_{\nu\mu}(s)\,.
\label{eq:j}
\eeqn
Note that due to the constraint~\eq{eq:constraint} the current~\eq{eq:j} is closed:
\beqn
\delta j \equiv \sum_{\mu}\partial'_{\mu}j_{\mu}(s) =
\sum_{\mu \neq \nu} \Bigl(n_{\mu\nu}^{(1)}(s)
-n_{\mu\nu}^{(1)}(s-\hat{\mu})+n_{\mu\nu}^{(2)}(s)
-n_{\mu\nu}^{(2)}(s-\hat{\nu}) \Bigr)=0\,.
\nonumber
\eeqn
Combining Eqs. \eq{eq:Fourier}, \eq{eq:Phi} and \eq{eq:j}, we get the {\it r.h.s} of
Eq.~\eq{eq:one} with
\beqn
\omega(j) & = & \sum_{\stackrel{n_{\mu\nu}^{(i)}\in{\Z}}{i=1,2,3}} \prod_s \Bigl[\prod_{i=1}^3
I_i(n_{\mu\nu}^{(i)}(s))\Bigr]\,
\delta\Bigl(j_{\mu}(s) - \sum_{\nu(<\mu)}n_{\mu\nu}^{(1)}(s) +\sum_{\nu(>\mu)}n^{(2)}_{\nu\mu}(s)\Bigr)
\nonumber\\
& \times &
\delta\Bigl(n_{\mu\nu}^{(1)}(s)-n_{\mu\nu}^{(1)}(s-\hat{\mu})- 2n_{\mu\nu}^{(3)}(s)\Bigr) \,
\delta\Bigl(n_{\mu\nu}^{(2)}(s)-n_{\mu\nu}^{(2)}(s-\hat{\nu})+2n_{\mu\nu}^{(3)}(s)\Bigr)\,.
\eeqn

\section{Schwinger-Dyson Equations}
\label{app:Schwinger:Dyson}

Consider a model of the gauge field $\theta$. The expectation value of an arbitrary
operator $O(\theta)$ measured at the ensemble $\{\theta_i\}$ of the gauge
fields $\theta$ is
\beqn
\langle O(\theta) \rangle = \int {\cal D}\theta \, O(\theta) \, e^{-S(\theta)} =
\prod_i \int\limits_{-\pi}^{\pi} d\theta_i \,
O(\{\theta_i\}) \hspace{1mm} e^{-S(\{\theta_i\})}\,.
\eeqn
Shifting one of the link fields $\theta_{i_0}$ at the link $i_0$ by an
infinitesimal value $\epsilon$ we get
\beqn
&&\prod_i \int\limits_{-\pi}^{\pi} d\theta_i \hspace{1mm}
O(\{\theta_i\}) \hspace{1mm}e^{-S(\{\theta_i\})}
= \prod_{i \ne i_0 } \int\limits_{-\pi}^{\pi} d\theta_i
\int\limits_{-\pi}^{\pi} d\theta_{i_0} \hspace{1mm}
O(\theta_{i_0},\{\theta_i\}_{i\ne i_0})
\hspace{1mm}e^{-S(\theta_{i_0},\{\theta_i\}_{i\ne i_0})}\nonumber\\
&\rightarrow&\prod_{i \ne i_0 } \int\limits_{-\pi}^{\pi} d\theta_i
\int\limits_{-\pi+\epsilon}^{\pi+\epsilon}
d\theta_{i_0} \hspace{1mm} O(\theta_{i_0}+\epsilon,\{\theta_i\}_{i\ne i_0})
\hspace{1mm}e^{-S(\theta_{i_0}+\epsilon,\{\theta_i\}_{i\ne i_0})}\nonumber\\
&=&\prod_{i \ne i_0 } \int\limits_{-\pi}^{\pi} d\theta_i
\int\limits_{-\pi}^{\pi} d\theta_{i_0} \hspace{1mm}
\Bigl( O(\theta_{i_0},\{\theta_i\}_{i\ne i_0})
\, e^{-S(\theta_{i_0},\{\theta_i\}_{i\ne i_0})}
\nonumber\\
&& +\epsilon \frac{\partial}{\partial \theta_{i_0}}
\Bigl[O(\theta_{i_0},\{\theta_i\}_{i\ne i_0})\hspace{1mm}
e^{-S(\theta_{i_0},\{\theta_i\}_{i\ne i_0})}\Bigr]+{\cal O}(\epsilon^2)\Bigr).
\eeqn
The requirement that this shift does not change the partition function gives the
Schwinger-Dyson equation:
\beqn
\prod_{i \ne i_0 } \int\limits_{-\pi}^{\pi} d\theta_i
\int\limits_{-\pi}^{\pi} d\theta_{i_0}
\frac{\partial}{\partial \theta_{i_0}}
\left[O(\theta_{i_0},\{\theta_i\}_{i\ne i_0}) \hspace{1mm}
e^{-S(\theta_{i_0},\{\theta_i\}_{i\ne i_0})}\right]
= \int {\cal D}\theta \frac{\partial}{\partial \theta_{i_0}}
\left[O(\theta) \, e^{-S(\theta)}\right]=0\,,
\label{eq:SD:0}
\eeqn
which can also be rewritten in the form:
\beqn
\left\langle\frac{\partial O(\theta)}{\partial \theta_{i_0}}\right\rangle
-\left\langle O(\theta) \frac{\partial S(\theta)}
{\partial \theta_{i_0}}\right\rangle=0\,.
\label{eq:SD}
\eeqn

To determine the parameters of the trial action \eq{eq:Seff:trial}-\eq{eq:s4}
we solve Eq.~\eq{eq:SD} with the following set of operators:
\begin{eqnarray}
O_I=\frac{\partial S_I}{\partial \theta_{\mu}(s)}
\hspace{3mm}(I=1,2,3,4),\hspace{5mm}
O_5=\frac{\partial S_4}{\partial \varphi_{\mu}(s)}.
\end{eqnarray}
The expectation values of these operators give
a set of five Schwinger-Dyson equations:
\begin{eqnarray}
\left \langle \frac{\partial^2 S_I}{\partial \theta_{\mu}(s)^2} \right \rangle
&=&\sum_{J=1}^3 \alpha_J \left\langle
\frac{\partial S_I}{\partial \theta_{\mu}(s)}
\frac{\partial S_J}{\partial \theta_{\mu}(s)}\right \rangle
+\beta_1\left\langle \frac{\partial S_4}{\partial \theta_{\mu}(s)}
\frac{\partial S_J}{\partial \theta_{\mu}(s)}\right \rangle
\hspace{5mm}(I=1,2,3,4),\label{eq:SQ:1}\\
\left \langle \frac{\partial^2 S_4}{\partial \varphi_{\mu}(s)^2}\right \rangle
&=&\beta_1\left\langle \left(
\frac{\partial S_4}{\partial \varphi_{\mu}(s)}\right)^2 \right\rangle.
\label{eq:SQ:2}
\end{eqnarray}

Since we have five equations \eq{eq:SQ:1}-\eq{eq:SQ:2} to determine
four independent couplings, $\alpha_i$, $i=1,2,3$ and $\beta_1$, the system
of equations \eq{eq:SQ:1}-\eq{eq:SQ:2} is overdefined. Thus we find
the couplings with the help of Eq.~\eq{eq:SQ:1} and then use
Eq.~\eq{eq:SQ:2} as a consistency check. We find that for the original
fields the {\it l.h.s.} of Eq.~\eq{eq:SQ:2} is approximately
10\% larger than the {\it r.h.s.} However, after applying the blockspin
transformation the discrepancy becomes much smaller (it becomes of the order of
the statistical errors), and the solution of Eqs.~(\ref{eq:SQ:1},\ref{eq:SQ:2})
becomes self--consistent.

\section{Hypercubic Blocking}
\label{app:Hypercubic:Blocking}

The Hypercubic Blocking (HYP) procedure is a version of the smearing method
which allows to reduce the noises of the lattice gauge fields~\cite{ref:HCB}.
As a result the statistical errors of ensemble averages of various operators
are reduced. HYP is replacing gauge link fields, $U_{\mu}(s)$, by "fat links",
$V_{\mu}(s)$, according to the following scheme:
\beqn
V_{\mu}(s)&=&\frac{1}{k_1}\left[(1-\alpha_1)U_{\mu}(s)
+\frac{\alpha_1}{6}\sum_{\nu\ne\pm\mu}\tilde{V}_{\nu;\mu}(s)
\tilde{V}_{\mu;\nu}(s+\hat{\nu})\tilde{V}_{\nu;\mu}^{\dag}(s+\hat{\mu})
\right],\nonumber\\
\tilde{V}_{\mu;\nu}(s)&=&\frac{1}{k_2}\left[(1-\alpha_2)U_{\mu}(s)
+\frac{\alpha_2}{4}\sum_{\rho\ne\pm\mu,\pm\nu}\bar{V}_{\rho;\mu\nu}(s)
\bar{V}_{\mu;\nu\rho}(s+\hat{\rho})\bar{V}_{\rho;\mu\nu}^{\dag}(s+\hat{\mu})
\right],
\label{eq:V-tilde}\\
\bar{V}_{\mu;\nu\rho}(s)&=&\frac{1}{k_3}\left[(1-\alpha_3)U_{\mu}(s)
+\frac{\alpha_3}{2}\sum_{\sigma\ne\pm\mu,\pm\nu,\pm\rho}U_{\sigma}(s)
U_{\mu}(s+\hat{\sigma})U_{\sigma}^{\dag}(s+\hat{\mu})
\right],\nonumber
\eeqn
where $k_i$, $i=1,2,3$ are chosen in such a way that the matrices~\eq{eq:V-tilde} belong to
the $SU(2)$ group. We choose the parameters of the HYP, $\alpha_1$, $\alpha_2$, $\alpha_3 \in [0,1]$,
following Ref.~\cite{ref:HCB}: $\alpha_1=0.75$, $\alpha_2=0.60$ and $ \alpha_3=0.30$.
At these values the smoothing of the gauge field configurations is most efficient.

\begin{acknowledgments}
We are thankful to Y.~Koma for generation of the vacuum configurations we used.
M.Ch. acknowledges the support by JSPS Fellowship No. P01023. T.S. is partially
supported by JSPS Grant-in-Aid for Scientific Research on Priority Areas
No.13135210 and (B) No.15340073. This work is also supported by the SX-5 at
Research Center for Nuclear Physics (RCNP) of Osaka University.
\end{acknowledgments}

\end{document}